\documentclass[superscriptaddress,floatfix,showkeys,pra]{revtex4-2}
\usepackage{amsmath,amssymb,amsthm,mathrsfs,amsfonts,dsfont,mathtools}
\usepackage[normalem]{ulem}
\usepackage{bbm}
\usepackage{physics}
\usepackage{graphicx,wasysym}   
\usepackage{xcolor}
\usepackage{float}
\usepackage[ruled,vlined]{algorithm2e}
\usepackage{tcolorbox}
\tcbuselibrary{theorems}
\usepackage{nicefrac}
\usepackage{enumerate}
\usepackage{hyperref}
\usepackage{newtxtext,newtxmath} 
\usepackage[section]{placeins}
\usepackage[caption=false]{subfig}
\usepackage{tabularx}

\newcommand{\fone}{{F1 }}
\newcommand{\sgn}[1]{\text{sgn}\left\lbrack#1\right\rbrack}
\newcommand{\Ket}[1]{\left|#1\right\rangle}
\newcommand{\Bra}[1]{\left\langle#1\right|}
\newcommand{\KetBra}[2]{\Ket{#1}\hspace{-3px}\Bra{#2}}
\newcommand{\BraKet}[2]{\left\langle#1 | #2\right\rangle}

\newtheorem{lemma}{Lemma}

\newtheorem{proposition}{Proposition}

\newtcbtheorem[number within=section]{note}{Note}%
{colback=blue!5,colframe=blue!35!black,fonttitle=\bfseries}{th}

\newcommand{\RR}{\mathbbm{R}}

\newcommand{\CC}{\mathbbm{C}}

\usepackage{xifthen}
\newcommand{\PP}[1][]{
  \ifthenelse{\isempty{#1}}
    {\mathbbm{P}}
    {\mathbbm{P}\left[#1\right]}
}
\newcommand{\EE}[1][]{
  \ifthenelse{\isempty{#1}}
    {\mathbbm{E}}
    {\mathbbm{E}\left[#1\right]}
}

\newcolumntype{P}[1]{>{\centering\arraybackslash}p{#1}}

\begin{document}
\title{Quantum-inspired classification via efficient simulation of Helstrom measurement}

\author{Wooseop Hwang}
\email{wooseop.hwang@exeter.ox.ac.uk}
\affiliation{Department of Physics, Oxford University, Oxford OX1 3PH, United Kingdom}
\author{Daniel K. Park}
\affiliation{Department of Applied Statistics, Yonsei University, Seoul 03722, Republic of Korea}
\affiliation{Department of Statistics and Data Science, Yonsei University, Seoul 03722, Republic of Korea}
\author{Israel F. Araujo}
\affiliation{Department of Statistics and Data Science, Yonsei University, Seoul 03722, Republic of Korea}
\author{Carsten Blank}
\email{blank@data-cybernetics.com }
\affiliation{Data Cybernetics, Landsberg 86899, Germany}

\date{\today}

\begin{abstract}
    The Helstrom measurement (HM) is known to be the optimal strategy for distinguishing non-orthogonal quantum states with minimum error. Previously, a binary classifier based on classical simulation of the HM has been proposed. It was observed that using multiple copies of the sample data reduced the classification error. Nevertheless, the exponential growth in simulation runtime hindered a comprehensive investigation of the relationship between the number of copies and classification performance. We present an efficient simulation method for an arbitrary number of copies by utilizing the relationship between HM and state fidelity. Our method reveals that the classification performance does not improve monotonically with the number of data copies. Instead, it needs to be treated as a hyperparameter subject to optimization, achievable only through the method proposed in this work. We present a Quantum-Inspired Machine Learning binary classifier with excellent performance, providing such empirical evidence by benchmarking on eight datasets and comparing it with 13 hyperparameter optimized standard classifiers.
\end{abstract}


\maketitle

\section{Introduction}
\label{section:introduction}

Quantum information (QI) theory introduces novel approaches and techniques for addressing various machine learning (ML) tasks with the potential to surpassing its classical counterparts~\cite{wittek,Biamonte_2017,Dunjko_2018,QML_PRSA,schuld2021machine,PhysRevLett.113.130503,lloydpca,RigorousRobustQSpeedUp, Huang_2022}. A salient quantum property pertinent to ML resides in the ability to distinguish non-orthogonal quantum states with a finite probability through quantum measurements~\cite{helstromoriginal,NC00,watrous_2018}. This unique attribute, which is absent in the classical theory, obviates the necessity to encode data with distinct labels as orthogonal states in quantum machine learning. As a result, quantum machine learning models can accommodate training datasets larger than the dimension of their associated Hilbert space. In this context, optimizing the success probability for distinguishing different states is of critical importance. This can be achieved via quantum state discrimination (QSD) techniques~\cite{Bae_2015}. Notably, the analytically derived set of measurements for optimal discrimination between two quantum states with minimal error is known as the Helstrom measurement, and the associated error is referred to as the Helstrom bound~\cite{helstromoriginal}. The theory of optimal QSD and the Helstrom measurement have catalyzed the development of various quantum machine learning (QML) algorithms. In the case of binary classification, it is the underlying principle of these QML algorithms, in that the measurement operators itself is determined by the training samples and is then applied on a test datum. This can be achieved either by running the algorithm on classical hardware~\cite{giuntini2021quantum,10.1007/s10773-017-3371-1,doi:10.1142/S0219749918400117,quantum-inspired}, termed quantum-inspired machine learning (QIML), or on quantum hardware via variational quantum algorithms~\cite{chen_qumi2020,PhysRevResearch.3.013063,Lee2023tsd}. Being able to run the algorithm on classical hardware is particularly interesting because it does not rely on the development of full-fledged quantum hardware, which remains to be a long-term prospect. However, computing the result of Helstrom's measurement is computational challenging on both classical and quantum hardware, arising from the need for full knowledge of the state and resource-use that grows exponentially. Consequently, its full exploitation within QML remains incomplete.

We investigate a particular QIML algorithm, the so-called Helstrom quantum centroid (HQC) classifier, which utilizes the Helstrom measurement to quantify the similarity of a data point with the two class centroids that are formulated as quantum states in the density matrix formulation~\cite{10.1038/s41534-020-0272-6,GIUNTINI2023109956, 10.1371/journal.pone.0216224}. There, it was reported that the performance can be substantially refined by conducting global measurements on a quantum state that encompasses multiple copies of the identical data-encoded state. The investigation into the effects of using multiple copies has been limited to a numerical study with a maximum of four copies, however, primarily due to the exponential growth of the memory and runtime as the number of copies increases. Therefore, it remained uncertain whether the classification performance increased monotonically with the number of copies, or if adverse effects, such as overfitting, started to impede performance as the data space continues to expand.

In this light, we report an efficient method for computing the HQC classifier on a classical computer, introducing a new (QIML) algorithm for distance-based binary classification. We call it the HQC-Simulation (HQCS) classifier, motivating its name from the fact that a modified version of the so-called fidelity (FID) classifier~\cite{10.1038/s41534-020-0272-6,PARK2020126422,lloyd2020quantum} is utilized to mimic the behavior of the HQC classifier. While the direct evaluation of Helstrom's measurement would require the computational and memory to grow exponentially with the number of copies, they remain constant in the HQCS classifier. This enables the investigation into the effects of the number of copies on classification performance at an arbitrary large scale.  Our findings reveal that the prediction performance of HQCS classifier does not increase monotonically with the number of copies, and the optimal number of copies varies depending on the dataset. This observation highlights the importance of optimizing the number of copies as a hyperparameter to achieve optimal classification performance. In contrast to its quantum meaning, our method treats the number of copies as a purely mathematical entity, thereby removing the constraint of being limited to integer values. 

The effectiveness of the approach is validated through numerical experiments, involving eight standard real-world datasets. The HQCS is compared against the previously mentioned FID classifier and 13 hyperparameter-optimized classical machine learning methods. The experimental results reveal that the optimal number of copies in both HQCS and the FID classifier often take non-integer values, and its effect on the prediction performance is not monotonic. This observation indicates that the classification performance of our proposed methods surpasses the capabilities of previous approaches by treating the number of copies as a hyperparameter (literally interpreting the “copies” parameter). Upon hyperparameter optimization, the HQCS demonstrates a slight but consistent advantage over the FID classifier. Furthermore, for specific datasets, the QIML algorithms are top-ranking among the classifiers tested in this work. It is noteworthy that the proposed algorithms hold particular interest from a practical perspective, as they do not rely on quantum hardware.

The remainder of the paper is organized as follows. We begin by setting up the problem and describing Helstrom measurement, and then introducing both the HQC and the FID classifiers. Subsequently, we present the propositions that facilitate the efficient simulation of the HQC for an arbitrary number of copies of the data-encoded quantum state, leading to the newly proposed QIML algorithm, the HQCS classifier. To verify the efficacy of our HQCS classifier, we compared its \fone score on real-world datasets with existing classical supervised learning models. Furthermore, we detail the supervised learning experiments conducted to test our strategy. This section provides an extensive analysis of the performance of the HQCS classifier compared to 14 existing algorithms, including the FID classifier, across eight real-world datasets. Finally, we discuss the relevance of the findings and future work. The methods section contains information on the numerical methods, and how we implemented the hyperparameter optimization, so that the reader can repeat our findings. 

\section*{Results}
\label{section:results}

\begin{figure*}[t]
    \centering
    \includegraphics[width=1\textwidth]{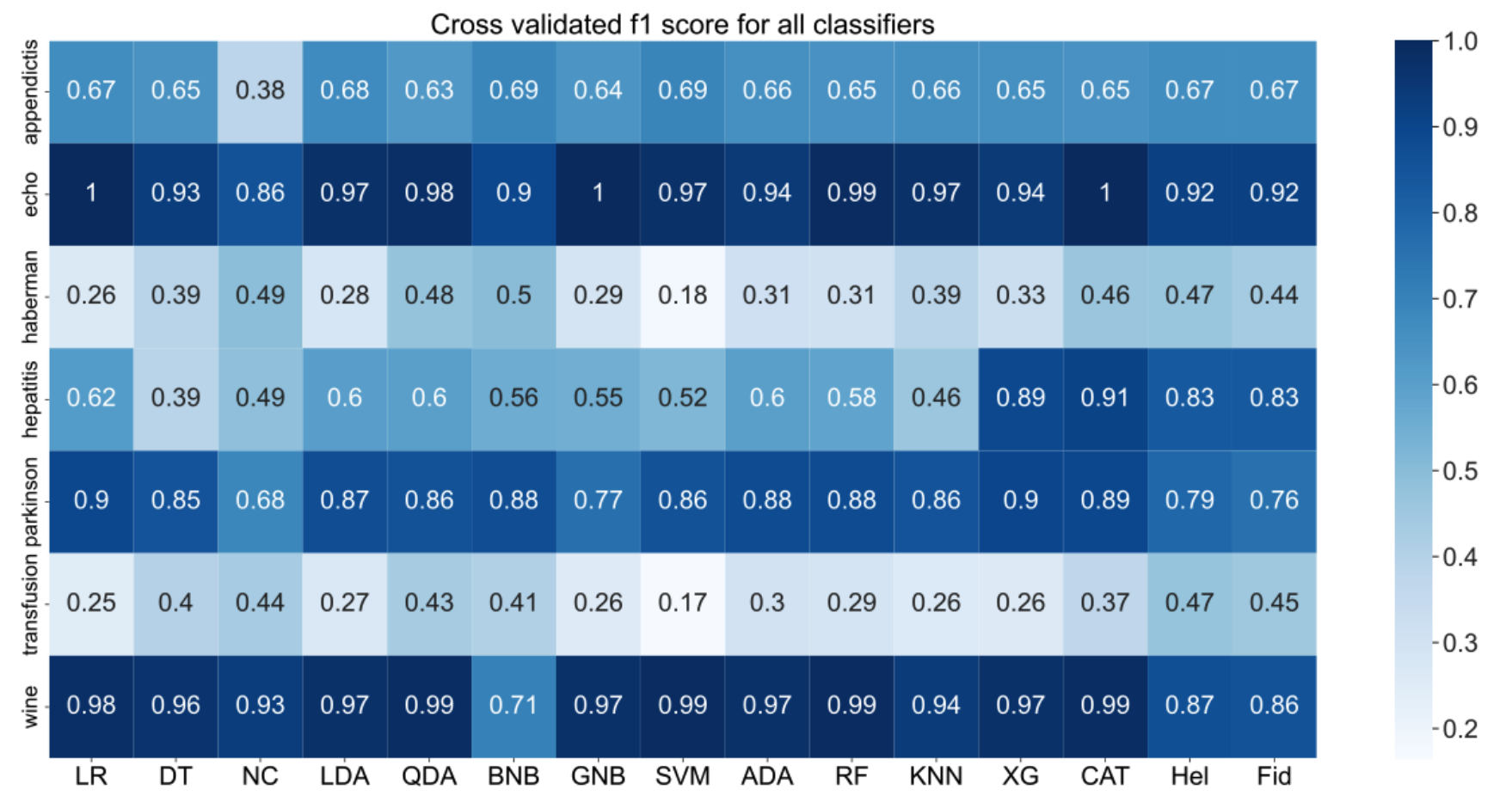}
    \caption{\textbf{Comparing \fone scores across different classifiers.} This figure illustrates the predictive performance of the classifiers across all datasets utilized in this study. Within the boxes, the \fone scores, derived from cross-validation, are presented. The classifier names are abbreviated for brevity, with full names provided in the main text. All classifiers were carefully fine-tuned in terms of hyperparameters using the ax-platform.}
    \label{fig:f1comparison}
\end{figure*}

This section defines the problem of binary classification and how to apply Helstrom's measurement to introduce both the HQC and the FID classifiers. The main result of this paper is then established, showing that the HQC classifier can be rewritten so that its evaluation takes a simple mathematical form, the HQCS. 

To give some context, the idea of the HQC classifier was first introduced by Sergioli et al. in Ref.~\cite{10.1371/journal.pone.0216224}. When encoding the data to quantum states into one quantum centroid per class, they explored the fact that multiple copies of a quantum state provided additional information compared to that encoded into a single copy, affecting classification performance~\cite{PhysRevLett.98.160501}. Despite this, the number of quantum copies they were able to explore was limited due to the exponential complexity required to perform Helstrom's measurement. For $k$ quantum copies and input length of $N$ numerical features, the previous computation involve $k$-fold Kronecker product resulting $O(N^k)$ for both memory usage and time complexity. As a result, the prediction performance of the HQC classifier could not be fully optimized, and its behavior at quantum copies greater than four remained unexplored. Through the identities used below, it is possible to explore this previously uncharted territory.    

\subsection*{Binary Classification \& Helstrom's Measurement}
    
Given two classes, the primary goal of a binary classifier is to determine a class in which an input data belongs to. In this work, we will denote these classes as class 0 and class 1. The framework of employing the Helstrom measurement to the binary classification tasks proceeds as follows. For classical binary classification tasks in the context of supervised learning, a labeled training dataset is usually given as real-valued vectors. To apply Helstrom's measurement, the classical data first has to be transformed to quantum states via a quantum feature map $\phi$. We identify for brevity the notations $\phi(x) = \Ket{\phi(x)} = \Ket{x}$ to describe the encoding of data $x\in\CC^d$ into the quantum Hilbert space, such that $\left|\BraKet{x}{x}\right|= 1$. There exists numerous schemes that can be implemented as feature maps such as amplitude encoding, angle encoding, and Hamiltonian encoding~\cite{PhysRevA.101.032308,supervised_featuremap,PhysRevA.102.032420,9259210,araujo_divide-and-conquer_2021,araujo2021configurable,Hur_2022,hur2023neural}. With a given feature map, we have a labelled training dataset in the proper Hilbert space: 
\begin{equation}
    D =  \{ (x_1, y_1), \dots, (x_M, y_M)\} \subseteq \mathbb{C}^d \times \{0,1\}.
\end{equation}
By denoting two subsets $D_0 = \{a: (a,0) \in D\}$ and $D_1 = \{b: (b,1) \in D\}$, we define quantum centroids with $k$ quantum copies as
\begin{align}
\label{centroid1}
\rho^{k} &= \frac{1}{M_a}\sum_{a} \left(\Ket{a}\Bra{a}\right)^{\otimes k} \\
\label{centroid2}
\sigma^{k} &= \frac{1}{M_b}\sum_{b} \left(\Ket{b}\Bra{b}\right)^{\otimes k},
\end{align}
where $M_a$ and $M_b$ represent the number of elements in $D_0$ and $D_1$, respectively. Note that the physical meaning of the quantum copies of a state $\Ket{\psi}$ is simply 
\begin{equation}
    \Ket{\psi}^{\otimes k} = \Ket{\psi} \otimes \Ket{\psi} \cdots \otimes \Ket{\psi},
\end{equation}
with $k$ number of identical replica of $\Ket{\psi}$. Under such conditions, the Helstrom operator is defined as
\begin{equation}
\label{eq:helstrom_operator}
    m_{\text{1}}^{k} = \rho^{k} - \sigma^{k}.
\end{equation}
This operator in equation~\eqref{eq:helstrom_operator} is Hermitian and can therefore be rewritten in terms of eigenbasis as
\begin{equation}
\label{eq:fidelity}
    m_{\text{1}}^{k} = \sum_{j} \lambda_{j,k} \Ket{d_{j,k}}\Bra{d_{j,k}},
\end{equation}
where $\lambda_{j,k} \in\RR$ and $\Ket{d_{j,k}}$ are the eigenvalues and eigenstates of $m_{\text{1}}$ respectively. Then, the Helstrom projection operators are given by~\cite{Bae_2015}
\begin{equation}
\begin{aligned}
    \Pi_{+} &= \sum_{\lambda_{j,k} \ge 0} \Ket{d_{j,k}}\Bra{d_{j,k}}\\
    \Pi_{-} &= \sum_{\lambda_{j,k} < 0} \Ket{d_{j,k}}\Bra{d_{j,k}}.
\end{aligned}
\end{equation}
$\Pi_{+}$ is responsible for projecting an input state to the eigenstates with non-negative eigenvalues, and $\Pi_{-}$ projects an input state to the eigenstates with negative eigenvalues. An analogous formulation as in equation~\eqref{eq:fidelity} can be formulated for the eigenbasis description
\begin{equation}
\label{eq:helstrom}
    m_{\text{2}}^k = \sum_{j} \sgn{\lambda_{j,k}}  \Ket{d_{j,k}}\Bra{d_{j,k}} = \Pi_{+} - \Pi_{-}.
\end{equation}    
For a given test data $c\in\CC^d$, HQC and fidelity classifiers are respectively defined as
\begin{align}
\label{eq:helstrom_classifier}
    f_{\text{hqc}}^k(c) &= \Bra{c}^{\otimes k} m_{\text{2}}^{k} \Ket{c}^{\otimes k} \\
\label{eq:fidelity_classifier}
    f_{\text{fid}}^k(c) &= \Bra{c}^{\otimes k} m_{\text{1}}^{k} \Ket{c}^{\otimes k}.
\end{align}    
Here it is visible why HQC has its naming: it defines quantum centroids and constructs Helstrom' measurement operator based on the two class quantum centroids. It is also apparent how the HQC and FID  classifiers relate to each other, if considered in their spectral decomposition, which plays a major role later. The outcomes of~\eqref{eq:helstrom_classifier} and~\eqref{eq:fidelity_classifier} are known as classification scores. The class of the test data $c$ is determined based on the sign of the classification scores as follows:
\begin{equation}
\label{eq:label_scheme}
   y_{\text{pred}} = 
    \begin{cases}
        0, & \text{if $f(c) > 0$} \\
        1, & \text{otherwise}
    \end{cases}
\end{equation}
From hereon, the term 'positive class' is used to refer to a set of data in class 0, while 'negative class' is used to indicate a set of data in class 1.

\subsection*{Theoretical Result}

The above equations show that a na\"ive computation of the classifiers involves the computation of eigenvalues and eigenbasis and that the copies increases the size of the problem---and thus the computational complexity---exponentially. In this section, we will show that the HQC can be identified with the FID by a non-linear transformation. To this end, we note that one important identity is true: 
\begin{equation}
\label{eq:helstrom_prob_identity}
    f^k_{\text{hqc}}(c) = \frac{1}{M_a M_b} \sum_a \sum_b f_{\text{hqc}}^{a,b,k}(c)
\end{equation}
where we used the notation $f_{\text{hel}}^{a,b,k}(c)$ to signify a single-datum per class train set $D_0 = \{a\}, D_1=\{b\}$. This identity is proven in the Appendix D of Lloyd et al.~\cite{lloyd2020quantum} and given in the Methods section of this work. The advantage of this identity lies in the fact that the function $f_{\text{hel}}^{a,b,k}(c)$ is easy to analyze: we need to know the eigenvalues of the Helstrom operator~\eqref{eq:helstrom_operator} and for the simplest case, where $D_0=\{a\}, D_1=\{b\}$, they can be computed analytically.
\begin{lemma}
\label{lemma:eigenvalues_m_2}
    Given two data points in opposing classes $a, b$ (and their encodings) and an integer $k>0$, the corresponding Helstrom operator $m_1^{a, b, k}$ is defined as
    \begin{equation}
    \label{eq:m_abk}
        m_1^{a, b, k} = (\Ket{a}\Bra{a})^{\otimes k} - (\Ket{b}\Bra{b})^{\otimes k}
    \end{equation}
    Then, the eigenvalues of the operator $m_1^{a, b, k}$ are given by $\pm \lambda_{a,b,k}$ with
    \begin{equation}
    \label{eq:eigenvalues}
        \lambda_{a,b,k} = \sqrt{1 - |\Bra{a}\Ket{b}|^{2k}}.
    \end{equation}
    and the operator is decomposed in
    \begin{equation}
        m_1^{a, b, k} = \lambda_{a,b,k}\left(\KetBra{d_{+,k}}{d_{+,k}} - \KetBra{d_{-,k}}{d_{-,k}}\right).
    \end{equation}
\end{lemma}    
The proof is also given in Methods and revolves around the examination of basis states and their coefficients in the given assumptions of the lemma. By using equations \eqref{eq:helstrom_classifier}, \eqref{eq:fidelity_classifier}, \eqref{eq:fidelity}, and \eqref{eq:helstrom}, we find ($\lambda \hat{=} \lambda_{a,b,k}$)
\begin{equation*}
\begin{aligned}
    &\frac{1}{\lambda}f_{\text{fid}}^{a,b,k}(c) \\
    &\quad= \lambda^{-1} \lambda \Bra{c}^{\otimes k} \Big(\KetBra{d_{+,k}}{d_{+,k}} - \KetBra{d_{-,k}}{d_{-,k}} \Big) \Ket{c}^{\otimes k} \\
    &\quad= \Bra{c}^{\otimes k} \Big( \sgn{\lambda} \KetBra{d_{+,k}}{d_{+,k}} + \sgn{-\lambda}\KetBra{d_{-,k}}{d_{-,k}} \Big) \Ket{c}^{\otimes k} \\
    &\quad= f_{\text{hqc}}^{a,b,k}(c).
\end{aligned}
\end{equation*}
Subsequently, by substituting the eigenvalues provided in equation~\eqref{eq:eigenvalues}, the following proposition naturally emerges.
\begin{proposition}
    Given two datasets in opposing classes $D_0, D_1$ (or their encodings) and an integer $k>0$, then we have
    \begin{equation}
    \label{eq:helstrom_as_fidelity_identity}
        f_{\text{hqc}}^{k} (c) = \frac{1}{M_a} \frac{1}{M_b} \sum_{a} \sum_{b} \frac{1}{\sqrt{1 - |\langle a|b \rangle|^{2k}}} f_{\text{fid}}^{a,b,k}(c)
    \end{equation}
    given with
    \begin{equation}
         f_{\text{fid}}^{a,b,k}(c) = |\langle c|a \rangle|^{2k} - |\langle c | b \rangle|^{2k}.
    \end{equation}
\end{proposition}
To summarize, by using the probabilistic aspect of quantum measurement, one can find a non-linear transformation that maps the expectation value of the Helstrom operator $m_2^k$ given the state $\Ket{c}$ to the evaluations of sums of inner products.

In Ref.~\cite{10.1371/journal.pone.0216224}, the quantum centroids (see equations \eqref{centroid1} and~\eqref{centroid2}) were computed by first computing the density matrix of each training data via $k$-fold Kronecker product, and then summing the $n$-fold density matrices and dividing them by the number of samples. The centroids are then used to construct the Helstrom operator~\eqref{eq:helstrom_operator}, which is necessary to run the FID classifier. The eigenstates of the Helstrom operator, which are essential for the construction of the HQC classifier, are identified through conventional eigen-decomposition employing the integrated Linear Algebra PACKage (LAPACK)\cite{lapack99}. It is the $k$-fold Kronecker product that gives rise to the memory-usage and time complexity of $O(N^k)$, given the input data length of $N$. 

By computing the eigenvalues of the Helstrom operator using the overlap between the data from different classes as shown in~\eqref{eq:eigenvalues}, the computation of $k$-fold Kronecker product is no longer required. As a result, the overall complexity of simulating the HQC classifier becomes $O(N)$. The number of quantum copies we can accurately investigate is still limited by the precision of the classical computer with the number of bits for representing the overlap, which becomes exponentially small with the number of copies. The standard 64-bit numeric precision can accurately represent numbers as small as $10 
^{-308}$. To handle values beyond this, higher precision methods such as bignum arithmetic are required. The effectiveness of such techniques is constrained by the available memory of a classical computer. Considering also encodings, the complexity of the computation of $|\langle a|b\rangle|$ depends on the type of quantum feature map used to encode classical data. When dealing with classically efficiently computable feature maps like amplitude and angle encodings, the overlap is calculated using a standard inner product. This approach exhibits linear complexity regarding the number of features and the amount of data. However, for feature maps that are challenging to simulate classically, quantum computers can be leveraged to compute the overlap. In such cases, techniques like the swap test~\cite{barenco1996stabilisation} and the inversion test~\cite{naturefeaturemap,lloyd2020quantum} can be executed on quantum computers to compute the overlap.

\subsection*{Benchmarking}
\label{section:numerical_experiments}

Having proposed the HQCS classifier, we show the benchmarking results comparing it with the FID classifier, and 13 standard classifiers: K-Nearest Neighbors (KNN), Linear Support Vector Classification (SVM), Decision Tree Classifier (DT), Random Forest Classifier (RF), Ada Boost Classifier (ADA), Bernoulli Naive Bayes Classifier (BNB), Gaussian Naive Bayes (GNB), Quadratic Discriminant Analysis (QDA), Linear Discriminant Analysis (LDA), Nearest Centroid Classifier (NC), Logistic Regression (LR), XGBoost~\cite{chen2016xgboost} (XGB) and CatBoost~\cite{prokhorenkova2018catboost} (CT). 

We investigated classification scores and prediction performances on eight different datasets: Appendicitis~\cite{romano2021pmlb}, Echo-cardiogram~\cite{misc_echocardiogram_38}, Hepatitis~\cite{misc_hepatitis_46}, Haberman ~\cite{ucidataset}, Iris~\cite{misc_iris_53}, Parkinson~\cite{misc_parkinsons_174}, Transfusion ~\cite{ucidataset} and Wine~\cite{misc_wine_109}. A five-fold cross-validation was performed on the entire data when computing the \fone scores and classification scores. There were no feature maps used, the features were treated directly as vectors. In the quantum computing realm, this is also called amplitude encoding~\cite{schuld2021machine}. Therefore, equation~\eqref{eq:helstrom_as_fidelity_identity} allowed for efficient simulation of the HQCS \& FID for large numbers of copies. 

As most of the datasets were unbalanced, the \fone score was used as the metric for comparing prediction performance. Figure~\ref{fig:f1comparison} presents the prediction performance comparison of the classical classifiers and the newly introduced classifiers over seven of the eight datasets. The \fone scores for the iris dataset are not displayed, as all classifiers achieved perfect classification scores. The results show that for datasets such as Echo-cardiogram and Parkinson, the HQCS and FID classifiers performed almost as well as the classical classifiers. On the other hand, the HQCS and FID classifiers exhibited a clear outperformance in comparison to hyperparameter-optimized classical binary classifiers for the Haberman, Hepatitis, and Transfusion datasets. This result highlights the potential of practical applications of the these quantum-inspired binary classifiers. This may suggest the existence of dataset structure that is more suitable for HQCS and FID classifiers than classical binary classifiers.

\section*{Discussion}
\label{section:conclusion}

Helstrom measurement is the optimal single-shot measurement that discriminates two arbitrary quantum states. The idea of incorporating the Helstrom measurement in binary classification was first proposed in ~\cite{10.1371/journal.pone.0216224}. We introduced a new classifier, the Helstrom-Quantum-Centroid-Simulation, which achieves the efficient simulation of the Helstrom-Quantum-Centroid algorithm on classical hardware and extends the number of quantum copies to an arbitrary positive value. These two achievements constitute the theoretical result of this work. Complementary, the experimental result is based on numerical experiments on real-world datasets from ~\cite{ucidataset} and ~\cite{romano2021pmlb}, where we show that the novel classifier, featuring only one hyperparameter, performs on par with 13 commonly used standard classifiers, consistently achieving competitive results even when compared to boosting techniques. Remarkably, the HQCS classifier is particularly strong on three datasets (Haberman, Hepatitis, and Transfusion).  One caveat exists. Despite the good performance, the proposed classifier belongs to the category of distance-based supervised learning algorithms and therefore does not conclude the training process within a model. Rather than extracting appropriate values based on training data, the distribution of all training data needs to be taken for classification of a test datum. When this is taken into account and is deemed appropriate with the application in question, we therefore present a solid new method for binary-distance-based classification.

The HQCS and fidelity classifier demonstrated comparable performance to classical classifiers, specifically on datasets like Appendicitis, Echocardiogram, Haberman, and Transfusion. We conducted a numerical analysis based on a method introduced in~\cite{Napierala} to assess the structure complexity of these datasets using the K-Nearest Neighbours (KNN) classifier. This method helps identify datasets with intricate structures, indicating closely-spaced or partially overlapping classes. Among the datasets studied, Appendicitis, Haberman, and Transfusion were deemed relatively challenging due to their complex structures. It's important to note that while a complicated structure doesn't guarantee poor prediction performance for every classifier~\cite{Napierala}, it does suggest that distance-based approaches might struggle. However, the HQCS and fidelity classifiers not only outperform distance-based classifiers like Nearest Centroid and KNN but also perform comparably to state-of-the-art classifiers like XGBoost and Catboost on these challenging datasets. This implies that the quantum-inspired distance-based binary classifiers introduced in this study excel in making accurate predictions even when classical distance-based approaches might falter. Incorporating quantum metrics into distance measures seems to enhance the classification of classically inseparable data, leading to improved prediction performance.

Interesting future work would be applying the kernel trick to theoretically compare the prediction performance of the HQCS and FID-classifier. The Helstrom measurement is known as the most ideal measurement in terms of minimizing the probability of error in two-state quantum discrimination~\cite{Bae_2015}, However, the optimal prediction performances of the HQCS and FID-classifier are the same for almost all datasets we have considered. The two classifiers obtain the most optimal performance at different copies. The Helstrom measurement is known to be the solution to lower error bound in the quantum state discrimination than the discrimination based on the fidelity. However, in the context of binary classification, Helstrom classifier may exhibit indistinguishable prediction performance when compared to the fidelity classifier. Implementing the kernel trick allows theoretical comparison, as the geometric distance between two kernels enables the theoretical comparison of the prediction performances of the kernels~\cite{Huang_2021}. We already calculated the associated kernel to the fidelity kernel, and it is presented in ~\ref{fidelity_kernel}. Investigation of the kernel for the HQC and HQCS classifier would be one of the promising directions of future research.

Another avenue for future exploration is to investigate quantum feature maps other than amplitude encoding. In particular, it would be worthwhile to investigate whether certain quantum feature maps that are hard to simulate classically can enhance the performance of the classifier for specific datasets~\cite{supervised_featuremap,RigorousRobustQSpeedUp}. In this case, a quantum hardware is necessary for computing the state overlap. Furthermore, recent techniques for finding the optimal quantum embedding for a given dataset~\cite{hur2023neural} may offer a promising approach for optimizing the performance of the classifiers proposed in this work.

Note that the Helstrom Measurement is limited to the discrimination of two different quantum states. Thus, the HQC and HQCS classifier is limited to binary classification tasks. For discrimination of more than two quantum states, Pretty Good Measurements (PGMs) are known as one of the best measurement schemes~\cite{PGMs}. There exists previous efforts to employ PGMs in the context of multi-class classification tasks~\cite{giuntini2021quantum}. The implementation of the kernel trick allowed the efficient simulation of PGMs for any number of copies~\cite{cruzeiro2023quantuminspired}. Given that the Helstrom Measurement is equivalent to the PGMs with two quantum states to discriminate, the investigation on the kernel of PGMs possess the potential of revealing the suitable kernel for the HQC and HQCS classifier.

\section*{Methods}

\subsection*{Formal proof of identity~\ref{eq:helstrom_prob_identity}}

A cornerstone of our method is the identity~\eqref{eq:helstrom_prob_identity} between two formulations of computing the expectation value of the Helstrom projector operator~\eqref{eq:helstrom} using its components. The proof is given in the Appendix D of~\cite{lloyd2020quantum} and is repeated here as a sketch for convenience.
\begin{proof}
It can be shown that there exists a quantum circuit that calculates Helstrom's measurement: given an initial state $\eta = \Ket{\psi}\Bra{\psi} \otimes \rho^k \otimes \sigma^k$ and a unitary operation $W$ and a measurement $M$, such that
$$
    \Tr{W\eta W^\dagger M} \stackrel{!}{=} \Tr{\KetBra{c}{c}^{\otimes k} m_1^k} = f_{\text{hel}}(c).
$$
These operators can be found, see~\cite{lloyd2020quantum}. We can then write 
$$
\begin{aligned}
    \eta &= \KetBra{\psi}{\psi} \otimes \rho^k \otimes \sigma^k 
    = \frac{1}{M_a M_b} \sum_a \sum_b \eta_{a,b} 
\end{aligned}
$$
for $\eta_{a,b} = \KetBra{\psi}{\psi} \otimes \KetBra{a}{a}^{\otimes k} \otimes \KetBra{b}{b}^{\otimes k}$ we find 
$$
    \Tr{W\eta W^\dagger M} = \frac{1}{M_aM_b} \sum_a \sum_b \Tr{W\eta_{a, b} W^\dagger M}
$$
which proves the identity.
\end{proof}

\subsection*{Proof of lemma~\ref{lemma:eigenvalues_m_2}}
\label{eigen proof}
The Helstrom operator is given in the equation \eqref{eq:helstrom_operator}. For $k$ quantum copies, it would be
\begin{equation}
\label{helstrom_op_k}
    m_1^{a,b,k} = \KetBra{a}{a}^{\otimes k} - \KetBra{b}{b}^{\otimes k},
\end{equation}
where $a$, and $b$ are elements from sets $D_0$ and $D_1$, respectively. Suppose $\Ket{d_j}$ is the $j$th eigenstate of the operator. We can express $\Ket{d_j}$ as follows.
\begin{equation}
    \Ket{d_{j}} = \sum_{i=1}^{2^k} c_{i} \Ket{i} + \Ket{\psi},
\end{equation}
where the $c_{i}$ are arbitrary constants, $\Ket{\psi}$ is an arbitrary state orthogonal to both $\Ket{a}$ and $\Ket{b}$. The states $\Ket{i}$ represent
\begin{gather*}
    \Ket{1} = \Ket{a} \otimes\Ket{a} \cdots \otimes \Ket{a} \otimes |a\rangle \\
    \Ket{2} = \Ket{a} \otimes \Ket{a} \cdots \otimes \Ket{a} \otimes \Ket{b} \\
    \vdots\\
    \Ket{2^k} = \Ket{b}\otimes\Ket{b} \cdots \otimes \Ket{b} \otimes \ket{b}.
\end{gather*}
for each of these states, there are $k$ copies of the combinations of $\Ket{a}$ and $\Ket{b}$. The eigenstates $\Ket{d_j}$ satisfy
\begin{equation}
\label{eigen_equation}
    m_1^{a,b,k} \Ket{d_j} = \lambda_{a,b} \Ket{d_j},
\end{equation}
where the $\lambda_{a,b}$ is an eigenvalue of $\Ket{d_j}$. The $m_1^{a,b,k}$ projects the input state to $|a\rangle^{\otimes k}$ and $|b\rangle^{\otimes k}$. LHS of the equation \ref{eigen_equation} would be
\begin{equation*}
\begin{aligned}
    m_1^{a,b,k}\Ket{d_j} = &\left[c_1 + (c_2 + c_{k-1})\BraKet{a}{b} + \cdots + c_k\BraKet{a}{b}^{k}\right]\Ket{a}^{\otimes k} \\
    - &\left[c_k + (c_2 + c_{k-1})\BraKet{a}{b} + \cdots + c_1\BraKet{a}{b}^{k}\right]\Ket{b}^{\otimes k}.
\end{aligned}
\end{equation*}
Equating this to the RHS immediately gives
\begin{equation*}
    c_i = 0, \quad \textrm{for } \quad i=2,\dots, 2^k-1.
\end{equation*}
We end up with two equations:
\begin{equation*}
\begin{aligned}
    c_1 + c_{2^k}\left|\BraKet{a}{b}\right|^{k} &= \lambda_{a,b} c_1 \\
    -c_{2^k} - c_1\left|\BraKet{a}{b}\right|^{k} &= \lambda_{a,b} c_{2^k},
\end{aligned}
\end{equation*}
Solving for $\lambda_{a,b}$, we obtain
\begin{equation}
    \lambda_{a,b} = \pm \sqrt{1 - \left|\BraKet{a}{b}\right|^{2k}}.
\label{eigenvalue_dervied}
\end{equation}
Note that this is an eigenvalue for a given $a$, and $b$.

\subsection*{Hyperparameter of the HQCS classifier}
In the following, we will explore the numerical experiment setup of both classifiers, the HQCS/FID-classifier. The number of copies of the system was the main parameter considered in the study. We now show that this parameter is actually a hyperparameter of the model. The classification scores are the outputs of the classifiers defined in~\eqref{eq:fidelity_classifier} and~\eqref{eq:helstrom_as_fidelity_identity}. The label determined by the classifier for test data, $y_{\text{pred}}$, was computed via~\eqref{eq:label_scheme}.  The \fone score was obtained by comparing $y_{\text{pred}}$ and the actual class for each test data $y_{\text{test}}$. We investigated the classification scores and \fone scores for number of quantum copies from 0.25 to 100 with step size of 0.25.

\begin{figure}[t]
     \centering
     \subfloat[Appendicitis]{%
        \includegraphics[width=0.5\textwidth]{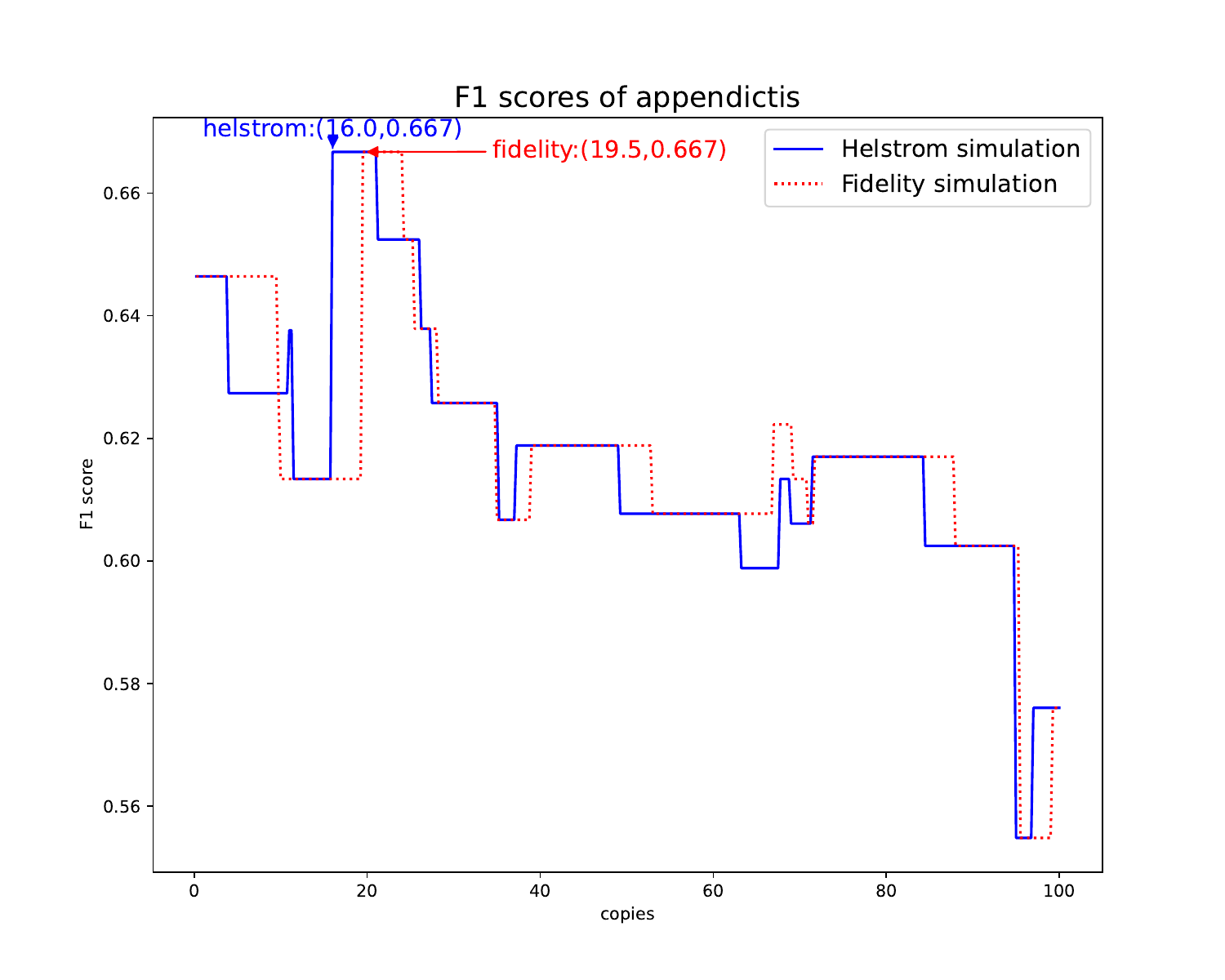} \label{fig:appendicitis f1}%
     }
     \subfloat[Transfusion]{%
        \includegraphics[width=0.5\textwidth]{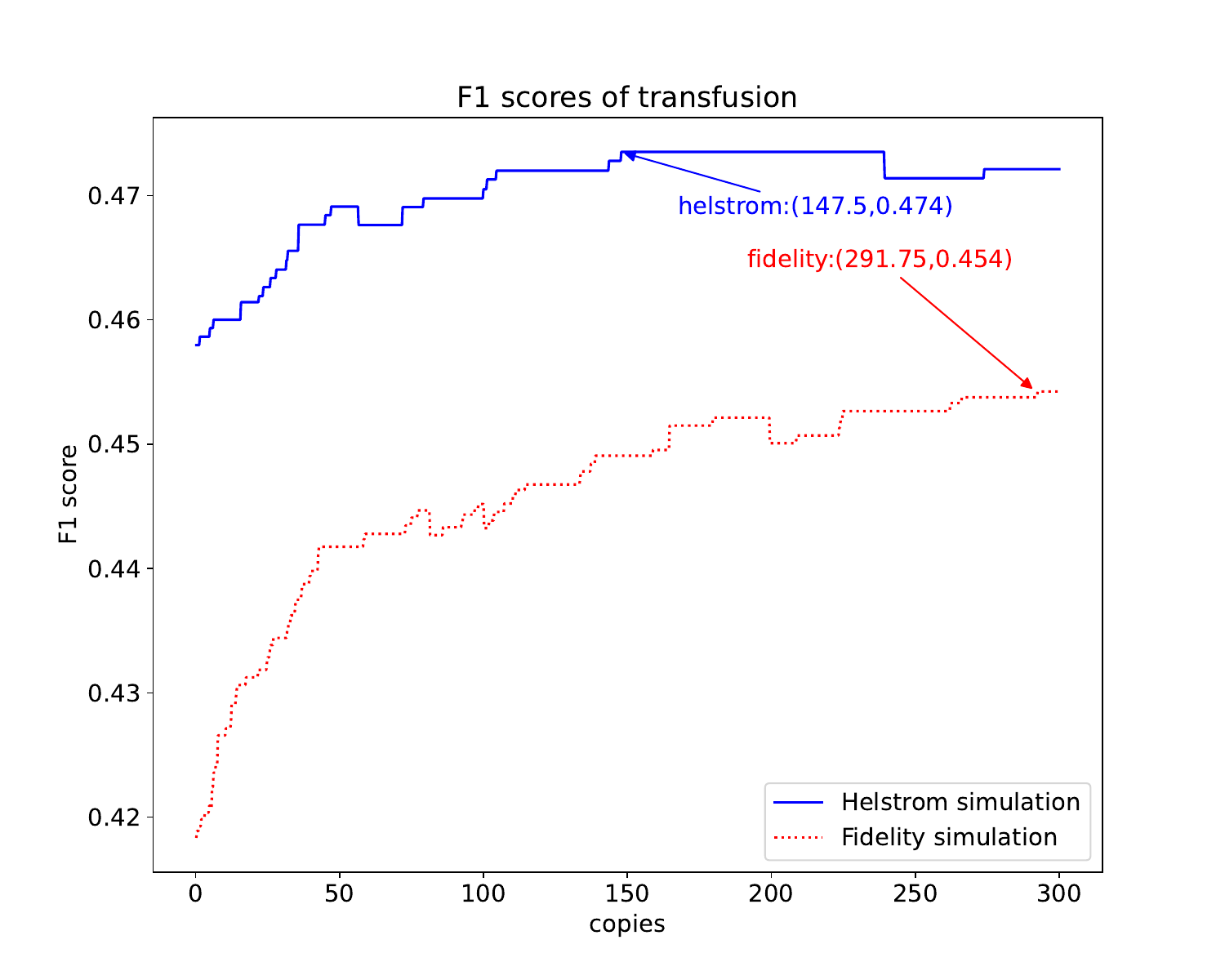} \label{fig:transfusion f1}%
     }
     \caption{\textbf{Cross validated \fone scores vs. parameter $k$}. \fone scores are shown in blue (HQCS) and red (FID). The \fone scores presented in the figures are the average of the five-fold cross validation. The maximum \fone scores, as well as the number of copies at which it occurs, are also annotated for each dataset. Results suggest non-monotonic impact of quantum copy count on prediction. HQCS+FID-classifiers perform comparably.}
     \label{fig:example_f1_scores}
\end{figure}

The simulation results for cross validated \fone score for the Appendicitis and the Transfusion datasets are displayed in Figure~\ref{fig:example_f1_scores}. Red and Blue were used to illustrate the \fone scores of the Helstrom classifier and the fidelity classifier, respectively. The annotations in Figure~\ref{fig:example_f1_scores} illustrate the number of copies required to achieve the maximum \fone score. Although the Helstrom bound has been conjectured to decrease as the number of copies increases~\cite{10.1371/journal.pone.0216224}, increasing the copies did not always lead to an improvement in the \fone score. Furthermore, while the HQCS/FID-classifiers tends to have the same maximum \fone scores for datasets, the maximum occurs at different number of copies for all datasets. These results indicate that the optimal number of copies is highly dependent on the dataset being analyzed, and therefore should be treated as hyperparameters. 
\subsection*{Hyperparameter Optimization}
For hyperparameter optimization of classical classifiers, we decided to use Ax (formerly known as AE~\cite{bakshy2018ae}). This platform uses Bayesian optimization to find appropriate parameters. 
\paragraph*{K-Nearest Neighbor (KNN)}
The hyperparameter search space we adopted for KNN is weights, number of neighbors, and distance metric. For weights, we chose between 'uniform' and 'distance'. For 'uniform', all points in the neighborhood were weighted uniformly, while for 'distance' points were weighted by the inverse of their distance. For number of neighbors, we chose an appropriate value in the range 1 to 10. For the distance metric, we chose between 'euclidean', 'minkowski', and 'manhattan'. 
\paragraph*{Support Vector Machine (SVM)}
For SVM, we considered regularization parameter ($C$), kernel, and kernel coefficient ($\gamma$) as hyperparameters. The search space of $C$ was in the range 0.1 to 1000. The kernels we considered were 'rbf', 'poly', 'sigmoid' and 'linear'. $\gamma$ was chosen in the range 0.1 to 1.0. 
\paragraph*{Decision Tree (DT)}
The hyperparameters we considered for DT were the criterion, the maximum depth, and the minimum number of samples required to split an internal node($min\_samples\_split$). For the criterion, we chose between '$gini$', '$entropy$', and '$log\_loss$'. The maximum depth was chosen in the range 1 to 10. The $min\_samples\_split$ was chosen in the range 2 to 10. 
\paragraph*{Random Forest (RF)}
The hyperparameters considered for RF are number of estimators, maximum depth, $min\_samples\_split$, and bootstrap. The number of estimator was chosen from a bound from 50 to 200. The maximum depth was chosen in the range 5 to 20. The $min\_samples\_split$ was chosen in range 2 to 10. For the bootstrap, we considered with and without the bootstrap. 
\paragraph*{Adaptive Boosting (ADA)}
The number of estimators, the learning rate and the algorithm were considered as hyperparameters for the ADA. The number of estimators were chosen in the range 1 to 200. The learning rate was chosen in between 0.1 and 1.0. The algorithms we considered were 'SAMME' and 'SAMME.R'. 
\paragraph*{Bernoulli Naive Bayes (BNB)}
For BNB, we considered the additive smoothing parameter ($\alpha$) and the threshold for binarizing of sample features(binarizing threshold) as hyperparameters. The search space for $\alpha$ was in the range 0 to 1.0, and that of binarizing threshold was in range 0 to 1.0 as well. 
\paragraph*{Quadratic Discriminant Analysis (QDA)}
For QDA, we set regularizing parameter as a hyperparameter and searched in the range 0 to 1. 
\paragraph*{Linear Discriminant Analysis (LDA)}
The hyperparameters considered for LDA were solver and shrinkage. We considered 'lsqa'and 'eigen' for solver and the shrinkage was chosen in the range 0.0 to 1.0. 
\paragraph*{Nearest Centroid (NC)}
For NC, we considered the distance metric, and shrink threshold as hyperparameters. The metrics we considered were 'minkowski', cityblock' and 'cosine'. The shrink threshold was chosen in the range 0 to 10. 
\paragraph*{Logistic Regression (LR)}
The hyperparameters considered for LR are the solvers, penalty, maximum iteration, and the inverse of regularization strength ($C$). The solvers we considered are 'lbfgs', 'liblinear', 'newton-cg', 'newton-cholesky', 'sag', and 'saga'. Appropriate penalty was chosen for each solvers, and maximum iteration was chosen from the range 100 to 1000. $C$ was chosen from the bound between 0.1 and 10.0. 
\paragraph*{XGBoost (XGB)}
For XGB, we set the learning rate, verbosity, and booster as hyperparameters. The learning rate was chosen from the range between 0.01 and 0.2. The verbosity were chosen from the  range between 0 and 3. The choices for booster was 'gbtree' and 'dart'. 
\paragraph*{Catboost (CAT)}
Lastly, the hyperparameters considered for the CAT were the learning rate, iterations, depth and the L2 regularization term of the cost function. The search space for the learning rate was in between 0.01 and 0.1. The iterations were chosen in the range 50 to 100. The depth was chosen from the range 6 to 12. The L2 regularization term was chosen from the bound 2 to 6.

\subsection*{Dataset categorization}
\label{datasetcategorization}

In this section, we introduce how the complexity of the dataset structure can be analyuzed through the use of KNN classifier. Most of the datasets considered in this work are all unbalanced, meaning that the distributions of the class labels are not equal. Compared to balanced data, they posses difficulty for classifiers in general~\cite{5128907, imbalanced, Chawla2005}. In such case, we can classify each test data into four distinct types: \textit{safe}, \textit{borderline}, \textit{rare} and \textit{outlier}, with increasing difficulty~\cite{Napierala}. In~\cite{Napierala}, authors present two different ways to classify each data: using k-nearest neighbours and implementing kernels. In this paper, we adopted the k-nearest neighbours approach. The approach focuses on the proportion of neighbours from the same class against neighbours from the opposite class with $k=5$ for each test data. Depending on this proportion, a given test data is classified as \textit{safe} (5:0 or 4:1), \textit{borderline} (3:2 or 2:3), \textit{rare} (1:4) and \textit{outlier} (0:5). To further categorize the entire dataset, we look at how the test data is distributed within \textit{safe}, \textit{borderline}, \textit{rare} and \textit{outlier}. As suggested in~\cite{Napierala}, Heterogeneous Value Difference Metric (HVDM)~\cite{hvdm} was employed as the metric for distance computation. We present the results in the table ~\ref{tab:difficulty}.

Interestingly, among all the datasets, Haberman had highest contributions from \textit{borderline}, \textit{rare}, and \textit{outlier} test data, indicating that it was the most difficult data for the learning models. Note that strong contribution from \textit{borderline}, \textit{rare}, and \textit{outlier} test data does not guarantee that all the classifiers will fail to make accurate predictions. Although the Hepatitis dataset possess small contribution of \textit{safe} data, the XGBoost and Catboost maintained high prediction performance \fone score of almost 0.9 as shown in Figure~\ref{fig:f1comparison}. However, the strong contributions from \textit{borderline}, \textit{rare}, and \textit{outlier} test data does indicate that distance-based classification will fail to perform well. Nearest Centroid classifier and KNN classifiers are the distance-based classifiers considered in this work, and we can observe that their poor prediction performance on datasets Haberman, Hepatitis, and Transfusion dataset which were categorized as difficult. Hence, we can conclude that classical distance-based classifiers are not likely to perfrom well on datasets with strong contributions from \textit{borderline}, \textit{rare}, and \textit{outlier} test data.

\begin{table}[h]
    \centering
    \begin{tabular}{|P{2.2cm}|P{1cm}|P{1cm}|P{1cm}|P{1cm}|}
        \hline
        Dataset & \textbf{S}(\%) & 
        \textbf{B}(\%) & \textbf{R}(\%) & \textbf{O}(\%) \\
        \hline
        Appendicitis & 49.6 & 
        17.1 & 5.7 & 27.6\\ 
        \hline
        Echocardiogram & 58.8 & 
        36.5 & 4.7 & 0 \\ 
        \hline
        Haberman & 4.9 & 
        39 & 34.3 & 21.7 \\ 
        \hline
        Hepatitis & 14.4 & 
        50.6 & 16.3 & 18.7 \\ 
        \hline
        Iris & 100 & 
        0 & 0 & 0 \\ 
        \hline
        Parkinson & 60 & 
        25.8 & 9.6 & 4.6 \\ 
        \hline
        Transfusion & 15.2 & 
        39.5 & 23.5 & 21.9 \\ 
        \hline
        Wine & 93.2 & 
        6.8 & 0 & 0 \\ 
        \hline

    \end{tabular}
    \caption{\textbf{Classification of datasets based on K-Nearest Neighbour approach introduced in ~\cite{Napierala}.}  We abbreviated the names of types, \textit{safe}, \textit{borderline}, \textit{rare}, and \textit{outlier}, with their first capital letter. Iris dataset solely consists of \textit{safe} test data, which explains why every classifier had perfect prediction performance. Among six datasets that were investigated in the main text, Haberman has the least contribution from \textit{safe} test data.}
    \label{tab:difficulty}
\end{table}

\subsection*{Kernel of the FID classifier}
In this section, we illustrate how the kernel of the FID classifier can be obtained. To express a classifier using kernel, we must find $w_{i}$ and $y_{i}$ such that~\cite{10.5555/1162264}

\begin{equation}
     f(c) = \sum_{i=1}^{N} w_{i} y_{i }K(c, x_i),
\end{equation}

where $K(c, x_i)$ is the kernel function, $x_i$ are the training data, and $c$ is the data we would like to classify. According to~\eqref{eq:fidelity_classifier}, the fidelity classifier with $k$ quantum copies is defined as 

\begin{equation}
    f_{\text{fid}}(c) = \frac{1}{M_a} \sum_a |\langle c|a\rangle|^{2k} - \frac{1}{M_b}\sum_b |\langle c|b\rangle|^{2k}.
\end{equation}

This can be rewritten as

\begin{equation}
    f_{\text{fid}}(c) = \sum_{i=1}^{N} w_i \left|\BraKet{c}{x_i}\right|^{2k}
\end{equation}

where 

\begin{equation*}
   w_{\text{i}} = 
    \begin{cases}
        \frac{1}{M_a}, & \text{for } y_i = 1 \\
        \frac{1}{M_b}, & \text{for } y_i = -1
    \end{cases}
.
\end{equation*}
\hfill

Hence, the suitable kernel function for the fidelity classifier would be
\begin{equation}
\label{fidelity_kernel}
    K(c,x_i) = \left|\BraKet{c}{x_i}\right|^{2k}.
\end{equation} Earlier in the paper, we have defined the relationship between the fidelity and the Helstrom classifier in~\eqref{eq:helstrom_as_fidelity_identity}. This equation, together with~\eqref{fidelity_kernel} does not lead to the helstrom kernel directly. The main reason for this is that the eigenvalues of the helstrom operator~\eqref{eq:eigenvalues} varies with the $x_i$, and is dependent on more than one input data.

\section*{Data Availability}

The site \url{https://github.com/seop02/Helstrom-simulation} contains all the data and the software generated during the current study.

\section*{Acknowledgments}
This research was supported by Institute for Information \& communications Technology Promotion (IITP) grant funded by the Korea government (No. 2019-0-00003, Research and Development of Core technologies for Programming, Running, Implementing and Validating of Fault-Tolerant Quantum Computing System), the Yonsei University Research Fund of 2023 (2023-22-0072), and the National Research Foundation of Korea (Grant Numbers: 2022M3E4A1074591, 2023M3K5A1094805, 2023M3K5A1094813), and the KIST Institutional Program (2E32941-24-008).

\section*{Conflict of interests}

C.B. is co-founder and managing director of data cybernetics ssc GmbH. All other authors declare no conflicts of interests.

\bibliographystyle{apsrev4-1}
\bibliography{main_v2} 

\end{document}